\documentclass[pra,aps,twocolumn,superscriptaddress,showpacs,showkeys]{revtex4}
\begin{document}
\author{Jing-Ling Chen}
\email{chenjl@nankai.edu.cn} \affiliation{Liuhui Center for Applied
Mathematics and Theoretical Physics Division, Chern Institute of
Mathematics, Nankai University, Tianjin 300071, People's Republic of
China}
\author{Kang Xue}
\affiliation{Department of Physics, Northeast Normal University,
 Changchun, Jilin 130024, People's Republic of China}
\author{Mo-Lin Ge}
\email{geml@nankai.edu.cn} \affiliation{Liuhui Center for Applied
Mathematics and Theoretical Physics Division, Chern Institute of
Mathematics, Nankai University, Tianjin 300071, People's Republic of
China}

\title {Berry phase and quantum criticality in Yang--Baxter systems}
\begin{abstract}
Spin interaction Hamiltonians are obtained from the unitary
Yang--Baxter $\breve{R}$-matrix. Based on which, we study Berry
phase and quantum criticality in the Yang--Baxter systems.
\end{abstract}

\pacs{03.67.Mn, 05.70.Jk, 03.65.Vf}

\keywords{Berry phase; quantum criticality; Yang--Baxter systems}
 \maketitle

\newcommand{\half}{{\textstyle \frac{1}{2}}}
\newcommand{\mn}{\mu\nu}
\newcommand{\eps}{\varepsilon}
\renewcommand{\L}{\mathcal{L}}

\newcommand{\eq}{\begin{equation}}
\newcommand{\en}{\end{equation}}
\newcommand{\eqa}{\begin{eqnarray}}
\newcommand{\ena}{\end{eqnarray}}

\def\l{\langle}
\def\r{\rangle}

\section{Introduction}
It is well known that the Yang--Baxter equation (YBE) plays a
fundamental role in the theory of $(1+1)$- or 2-dimensional
integrable quantum systems, including lattice statistical models and
nonlinear field theory. The YBE was originated in solving the
$\delta$-function interaction model by Yang \cite{YangCN} and the
statistical models by Baxter \cite{Baxter} and introduced to solve
many quantum integrable models by Faddeev and Leningrad Scholars
\cite{Faddeev}. Through the RTT relation \cite{Jimbo1} the new
algebraic structures (quantum groups) were established by V.
Drinfeld \cite{Drinfeld}. The usual YBE takes the form

\begin{equation}\label{1.1}
\breve{R}_{12}(x)\breve{R}_{23}(xy)\breve{R}_{12}(y)=\breve{R}_{23}(y)\breve{R}_{12}(xy)\breve{R}_{23}(x),
\end{equation}
which is valid for three types of $\breve{R}$-matrices, i.e., the
rational, the trigonometric and the elliptic solutions of YBE. The
spectral parameter $x$ plays an important role that is related the
one-dimensional momentum (or the rapidity) in some typical models.
An alternatively equivalent form of the YBE reads
\begin{equation}\label{1.1a}
\breve{R}_{12}(u)\breve{R}_{23}(u+v)\breve{R}_{12}(v)=\breve{R}_{23}(v)\breve{R}_{12}(u+v)\breve{R}_{23}(u),
\end{equation}
if one denotes $x=e^u$ and $y=e^v$ (or $x=e^{iu }$ and $y=e^{iv}$).
The asymptotic behavior of $\breve{R}_{i,i+1}(x)$ is
$x$-independent:
\begin{equation}\label{1.2a}
\lim\breve{R}_{i,i+1}(x)=B_{i}.
\end{equation}
The braiding operators $B_{i}$'s satisfy the following braid
relations

\begin{eqnarray} \label{braid}
   B_{i}  B_{i+1} B_{i} &=& B_{i+1} B_{i} B_{i+1}, \qquad
   1 \leq i \leq n-1, \nonumber\\
   B_{i} B_{j} &=& B_{j} B_{i}, \qquad\qquad\;\;\;
   | i - j | \ge 2,
\end{eqnarray}
where the notation $B_i \equiv B_{i,i+1}$ is used, $B_{i,i+1}$
implies ${\bf 1}_1 \otimes {\bf 1}_2 \otimes {\bf 1}_3 \cdots
\otimes B_{i,i+1} \otimes \cdots \otimes {\bf 1}_n$, and ${\bf 1}_j$
represents the unit matrix of the $j$-th particle. The usual
permutation operator
\begin{equation} \label{permutation} P_{i,i+1} =
\frac{1}{2}(1+{\vec \sigma}_i\cdot {\vec \sigma}_{i+1}) =
 \left( \begin{array}{cccc} 1 & 0 & 0 & 0 \\ 0 & 0 & 1 & 0 \\
0 & 1 & 0 & 0 \\ 0 & 0 & 0 & 1 \end{array} \right)
\end{equation}
 for the
particles $i$ and $i+1$ is a solution of Eq. (\ref{braid}) with the
constraint $P_{i,i+1}^2 =
 1$, where ${\vec \sigma}$ is the vector of the Pauli matrices.
 The permutation operator $P_{i,i+1}$ exchanges the spin state
$|k\rangle_i\otimes|l\rangle_{i+1}$ to be
$|l\rangle_i\otimes|k\rangle_{i+1}$.


For a statistical model all the elements of $\breve{R}(u)$-matrix
should be positive because they are related to the Boltzmann
weights. The relationship between $\breve{R}(u)$ and $B$ was set up
by Jimbo \cite{Jimbo2}, Jones \cite{Jones} and others
\cite{GeWuXue}. We call the process of obtaining the matrix
$\breve{R}(u)$ from a given braiding matrix $B$ as the
``\emph{Yang-Baxterization}'', which depends on the number of the
distinct eigenvalues of matrix $B$. As was pointed out by Kauffman
\emph{et al.} \cite{Kauffman}\cite{Kauffman1} that the braiding
matrix $B^{\frac{1}{2}\frac{1}{2}}$ (here the superscript
``$\frac{1}{2}\frac{1}{2}$'' means that the spin values of two
particles are both $\frac{1}{2}$) transforms the ``natural basis"
($\left|\uparrow\uparrow\right>$, $\left|\uparrow\downarrow\right>$
$\left|\downarrow\uparrow\right>$,
$\left|\downarrow\downarrow\right>$) to the Bell states
($\frac{1}{\sqrt{2}}\left(\left|\downarrow\downarrow\right>\pm\left|\uparrow\uparrow\right>\right)$,
$\frac{1}{\sqrt{2}}\left(\left|\downarrow\uparrow\right>\pm\left|\uparrow\downarrow\right>\right)$)
. It is emphasized that the elements of $B^{\frac{1}{2}\frac{1}{2}}$
are no longer positive here. However, a braiding matrix $B$ is
nothing to do with the dynamics. To do so, we should Yang-Baxterize
the matrix $B$ to be the $\breve{R}(x)$-matrix and look for its
resultant consequence in physics, such as Berry phase, quantum
criticality and so on.

In this paper, we mainly focus on the trigonometric
Yang-Baxterization. The rational Yang-Baxterization will be also
discussed, but the elliptic solutions of YBE will be ignored since
it is more complicated. If $B$ has only two distinct non-zero
eigenvalues $\lambda_1$ and $\lambda_2$, one then simply has the
trigonometric Yang-Baxterization approach
\cite{Jones,GeWuXue,Ge1994} as
\begin{equation}\label{1.4a}
\breve{R}(x)=\rho(x)\; (\lambda_1 x B+ \lambda_2^{-1} x^{-1}B^{-1}),
\end{equation}
where $\rho(x)$ is a normalization factor (one may choose an
appropriate $\rho(x)$ so that $\breve{R}(x)$ becomes a unitary
matrix).
Generally a solution of $\breve{R}(x)$ depends on two parameters:
the first is $\theta$ (or $x$, which is a function of $\theta$); the
second is $\phi$ contained in the braiding matrix $B$ (the free
parameter $\phi$ may be originated from the $q$-deformation
parameter with $q=e^{i \phi}$, or from other parameter such as
$\eta=e^{i \phi}$). In physics the parameter $\phi$ is the flux that
can be dependent on time $t$. Usually one takes $\phi=\omega t$ and
$\omega$ is the frequency. Alternatively, we can rewrite
$\breve{R}(x)$ as $\breve{R}(\theta, \phi(t))$.

Indeed, the YBE is a kind of fruitful resource that may provide
multi-spin interaction Hamiltonians. If we define a quantum state
through
\begin{equation}\label{1.2}
|\Phi(\theta, \phi (t))\rangle = \breve{R}(\theta, \phi(t))|\Phi(0)
\rangle,
\end{equation}
where $|\Phi(0) \rangle$ is the initial state independent of $t$,
and $\theta$ is time-independent. The normalization condition of the
quantum states
$\langle\Phi(\theta,\phi(t))|\Phi(\theta,\phi(t))\rangle=\langle\Phi(0)|\Phi(0)\rangle=1$
requires the unitary condition
$\breve{R}^{\dag}(\theta,\phi(t))=\breve{R}^{-1}(\theta,\phi(t))$.
 It follows from Eq. (\ref{1.2}) that
\begin{equation}\label{1.4}
\begin{array}{ll}
&i\hbar\frac{\partial|\Phi(\theta,\phi(t)\rangle}{\partial
t}\\
&=i\hbar\left[\frac{\partial\breve{R}(\theta,\phi(t))}{\partial
t}\breve{R}^{\dag}(\theta,\phi(t))\right]\breve{R}(\theta,\phi(t))|\Phi(0)\rangle\\
&=H(t)|\Phi(\theta,\phi(t))\rangle,
\end{array}
\end{equation}
where the Hamiltonian reads
\begin{equation}\label{1.5}
H(t)=i\hbar\frac{\partial\breve{R}(\theta,\phi(t))}{\partial
t}\breve{R}^{\dag}(\theta,\phi(t)).
\end{equation}
Thus, through the Yang-Baxterization approach
$B(\phi)\rightarrow\breve{R}(\theta,\phi)$,
Eq. (\ref{1.5}) defines the Hamiltonian for the Yang--Baxter
systems.


The purpose of this paper is to investigate some physical
consequences such as Berry phases (BP) \cite{Berry} \cite{Wilczek}
in Yang--Baxter systems, quantum criticality (QC) phenomenon
\cite{Fazio,Vidal,Wu,Vedral,Birman} is also discussed. The paper is
organized as follows. In Sec. II we study the Berry phase for a kind
of Yang--Baxter Hamiltonian related to the extra-special two-group.
In Sec. III, we study Berry phase for a Yang--Baxter Hamiltonian
related to the well-known six-vertex model and the Temperley-Lieb
(TL) algebra. Conclusion and discussion are made in the last
section.

\section{BP and QC for Hamiltonian $H_1(\theta,\phi(t))$}


Let us consider the following type braiding matrix for two spin-1/2
particles \cite{Kauffman}\cite{Kauffman1}

\begin{equation}\label{1.7}
B^{\frac{1}{2}\frac{1}{2}}=\frac{1}{\sqrt{2}}(I+M^{\frac{1}{2}\frac{1}{2}}),
\end{equation}
where $I$ is the $4 \times 4$ unit matrix,
\begin{equation}\label{1.8}
M^{\frac{1}{2}\frac{1}{2}}=\left(
\begin{array}{cccc}
&&&e^{i\phi}\\
&&\epsilon&\\
&-\epsilon&&\\
-e^{-i\phi}&&&
\end{array}\right),
\end{equation}
$\epsilon=\pm 1$, and $\phi=\phi(t)$ represents the arbitrary flux.
The braiding matrix $B^{\frac{1}{2}\frac{1}{2}}$ has special
significance in quantum information and quantum computation, because
it can be identified to the universal quantum gate (i.e., the CNOT
gate) \cite{Kauffman}\cite{Kauffman1}. In additional, the braiding
matrix $B^{\frac{1}{2}\frac{1}{2}}$ may produce the maximally
entangled states (or the Bell states) from the separable ones
$\left|\uparrow\uparrow\right>$, $\left|\uparrow\downarrow\right>$
$\left|\downarrow\uparrow\right>$, and
$\left|\downarrow\downarrow\right>$
\cite{Kauffman}\cite{Kauffman1}\cite{Chen2007}.

Furthermore, the matrix $M^{\frac{1}{2}\frac{1}{2}}$ satisfies the
algebraic relation of the extra-special two-group
\cite{Franko}\cite{Zhang2007}\cite{Fulton}\cite{Griess}. More
explicitly, the matrices $M_i^{\frac{1}{2}\frac{1}{2}}$ and
$M_j^{\frac{1}{2}\frac{1}{2}}$ satisfy the following algebraic
relations:

\begin{eqnarray} \label{extra}
&&M_i^2=-1,\nonumber\\
&& M_i M_j=M_j M_i,\,\, |i-j|\geq 2, \nonumber\\
&&M_{i+1}M_i=-M_i M_{i+1}, \,\, 1\leq i,j\leq n-1. \end{eqnarray} It
is easy to verified that the braiding matrix
$B^{\frac{1}{2}\frac{1}{2}}$ has two distinct eigenvalues with
$\lambda_1=(1+i)/\sqrt{2}$, $\lambda_2=(1-i)/\sqrt{2}$ and
$\lambda_1 \lambda_2=1$, then the trigonometric Yang-Baxterization
approach is applicable. Based on which one obtains
\begin{eqnarray*}
\breve{R}(x)&=&[2(x^2+x^{-2})]^{-1/2}[(x+x^{-1})I^{\frac{1}{2}\frac{1}{2}}\nonumber\\
&&+(x-x^{-1})M^{\frac{1}{2}\frac{1}{2}}],
\end{eqnarray*}
\begin{eqnarray}
[\breve{R}(x)]^{-1}&=&[2(x^2+x^{-2})]^{-1/2}[(x+x^{-1})I^{\frac{1}{2}\frac{1}{2}}\nonumber\\
&&-(x-x^{-1})M^{\frac{1}{2}\frac{1}{2}}]. \label{1.9}
\end{eqnarray}
The unitary condition $[\breve{R}(x)]^{-1}=\breve{R}(x^{-1})$ leads
to $\phi(t)=$real. In addition, the Yang--Baxter matrix
$\breve{R}(x)$ may produce the non-maximally entangled states when
it acts on the separable ones $\left|\uparrow\uparrow\right>$,
$\left|\uparrow\downarrow\right>$ $\left|\downarrow\uparrow\right>$,
and $\left|\downarrow\downarrow\right>$
\cite{Kauffman}\cite{Kauffman1}\cite{Chen2007}.

Equation (\ref{1.8}) can be rewritten as
\begin{eqnarray}\label{1.10}
M^{\frac{1}{2}\frac{1}{2}}&=&e^{i\phi(t)}S_1^+S_2^+-e^{-i\phi}S_1^-S_2^-\nonumber\\
&&+\epsilon(S_1^+S_2^--S_1^-S_2^+),
\end{eqnarray}
where
\begin{eqnarray}
S^+_i=S_i^1+ iS_i^2=\left(
\begin{array}{cc}
0&1\\
0&0
\end{array}
\right)_i, \nonumber\\
  S^-_i=S_i^1- iS_i^2=\left(
\begin{array}{cc}
0&0\\
1&0
\end{array}
\right)_i,
\end{eqnarray}
are the raising and lowering operators of spin-1/2 angular momentum
for the $i$-th particle, respectively. We then have from Eq.
(\ref{1.5}) that
\begin{eqnarray}\label{1.11}
H_1(x,\phi(t))&&=-\hbar\dot{\phi}[2(x^2+x^{-2})]^{-1}(x-x^{-1})
\times \nonumber\\
&&\{(x-x^{-1})(S_1^3+S_2^3)+\nonumber\\
&&(x+x^{-1})(e^{i\phi}S_1^+S_2^++e^{-i\phi}S_1^-S_2^-)\}.
\end{eqnarray}
By using
\begin{eqnarray}\label{1.12}
&&x=[-\cos2\theta]^{-1/2}(\cos\theta+\sin\theta),\nonumber\\
&&x^{-1}=[-\cos2\theta]^{-1/2}(\sin\theta-\cos\theta),
\end{eqnarray}
Equation (\ref{1.11}) can be recast to
\begin{eqnarray}\label{1.14}
H_1(\theta,\phi(t))&=&-\hbar\dot{\phi}\cos\theta[\cos\theta
(S_1^3+S_2^3)+\sin\theta(e^{i\phi}S_1^+S_2^+\nonumber\\
&&+e^{-i\phi}S_1^-S_2^-)],
\end{eqnarray}
or in the matrix-form it reads
\begin{eqnarray} \label{hamiltonian1}
H_1(\theta,\phi)&=& -\hbar {\dot \phi} \cos\theta
\left(\begin{array}{cccc}
\cos\theta & 0 & 0 & e^{i\phi}\sin\theta \\
0 & 0&  0 & 0 \\
0 & 0 & 0 & 0 \\
e^{-i\phi}\sin\theta& 0 & 0 & -\cos\theta \end{array} \right).
\end{eqnarray}

The eigen-problem of Eq. (\ref{1.14}) under
adiabatic approximation is
\begin{equation}\label{1.15}
H_1(\theta,\phi(t))|\Phi_\pm(\theta,\phi(t))\rangle_1=E_\pm^1(t)|\Phi_\pm(\theta,\phi(t))\rangle_1,
\end{equation}
where the two non-zero eigenvalues are
\begin{eqnarray}\label{1.16}
E_\pm^1&=&\mp\hbar\dot{\phi}\cos\theta
\nonumber\\
&=&\mp\hbar\omega\cos\theta\ \ \ \textrm{for}\ \phi=\omega t,
\end{eqnarray}
and the corresponding eigenstates are
\begin{equation}\label{1.17}
\begin{array}{rcl}
|\Phi_+(\theta,\phi)\rangle&=&\cos\frac{\theta}{2}|\uparrow\uparrow\rangle+
\sin\frac{\theta}{2}e^{-i\phi}|\downarrow\downarrow\rangle,\\
|\Phi_-(\theta,\phi)\rangle&=&-\sin\frac{\theta}{2}e^{i\phi}|\uparrow\uparrow\rangle+
\cos\frac{\theta}{2}|\downarrow\downarrow\rangle.
\end{array}
\end{equation}
The physical consequence of Berry phase for the above Yang--Baxter
Hamiltonian system, i.e., $H_1(\theta,\phi(t))$, has been discussed
in \cite{Chen2007}.
Namely, from the definition of Berry phase
\begin{eqnarray}\label{1.19}
\gamma(c)&=&i\int_0^Tdt\langle n(\vec{R})|\frac{\partial}{\partial
t}|n(\vec{R})\rangle =i\int_0^Tdt A(t)\nonumber\\
&=&i\int_0^{2\pi}dt\dot{\phi}^{-1}\langle
n(\vec{R})|\frac{\partial}{\partial \phi}|n(\vec{R})\rangle,
\end{eqnarray}
here $\vec{R}=(\sin\theta\cos\phi,\sin\theta\sin\phi,\cos\theta)$
and $|n(\vec{R})\rangle=|\Phi_{\pm}(\theta,\phi)\rangle$, one then
obtains the Berry phases for the Yang--Baxter system as
\begin{equation}\label{1.20}
\gamma_{\pm}^1=(\pm\int_0^{2\pi}d\phi)\sin^2\frac{\theta}{2}=\pm\pi(1-\cos\theta)=\pm\frac{\Omega}{2},
\end{equation}
where $\Omega=2\pi(1-\cos\theta)$ is the familiar solid angle
enclosed by the loop on the Bloch sphere.

The Hamiltonian (\ref{1.14}) is obtained through the Schr\"odinger
evolution of the Bell state with $\phi=\omega t$,
%
which does have a nice physical interpretation. Since ${\bf S}_1$
and ${\bf S}_2$ are two-dimensional representation operators of
$SU(2)$ for particles $1$ and $2$ respectively, we then have
$(S^{\pm}_i)^2=0$, $(i=1,2)$. It is convenient to introduce the
following fermionic operators:
\begin{equation}\label{2.2}
S^-_i=\left(
\begin{array}{cc}
0&0\\
1&0
\end{array}
\right)_i=\hat{f}_i,\ \ \ \ \ \ \ S^+_i=\left(
\begin{array}{cc}
0&1\\
0&0
\end{array}
\right)_i=\hat{f}_i^{\dag}
\end{equation}
Then by means of $[S^3_i, S^\pm_j]=\pm S^\pm_i\delta_{ij}$ and
$[S^+_i, S^-_j]=2\delta_{ij}S^3_i$, we have
\begin{eqnarray}\label{2.3}
&&\{\hat{f}_i,\hat{f}_i^\dag \}= \hat{f}_i\hat{f}_i^\dag
+\hat{f}_i^\dag \hat{f}_i=1, \nonumber\\
&&[\hat{f}_i,\hat{f}_j]=[\hat{f}_i,\hat{f}_j^\dag]=0, \;\;
(\textrm{for}\;\; i\ne j ), \nonumber\\
&&S^3_i=\hat{f}^{\dag}_i\hat{f}_i-\frac{1}{2}=\hat{n}_i-\frac{1}{2}.
\end{eqnarray}
i.e., $\hat{f}_i$'s satisfy the fermonic anticommutator for the same
$i$-th lattice and the bonsonic commutator for different sites of
the lattices, and $\hat{n}_i=\hat{f}^{\dag}_i\hat{f}_i$ is the
number operator that can be 0 and 1. It is easy to check that the
following three operators
\begin{eqnarray}\label{2.3a}
&& S^+=\prod^2_{i=1}\hat{f}^{\dag}_i,\;\;\;
S^{-}=\prod^2_{i=1}\hat{f}_i,\nonumber\\
&&S^3=\frac{1}{2}(S^{3}_{1}+S^{3}_{2})=\frac{1}{2}(\hat{n}_1+\hat{n}_2-1),
\end{eqnarray}
form an $SU(2)$ group satisfying $[S^3, S^\pm]=\pm S^\pm$, and
$[S^+, S^-]=2 S^3$. By the way, its Casimir operator is
$\frac{1}{2}(S^+S^-
+S^-S^+)+(S^3)^2=\frac{3}{4}[\hat{n}_1\hat{n}_2+(1-\hat{n}_1)(1-\hat{n}_2)]
=\frac{3}{4}[1-(\hat{n}_1-\hat{n}_2)^2]$, which equals to
$\frac{1}{2}(\frac{1}{2} + 1)$ for $\hat{n}_1=\hat{n}_2$, and
$0(0+1)$ for $\hat{n}_1 \ne \hat{n}_2$, respectively.

In terms of Eqs. (\ref{2.2})-(\ref{2.3a}) the Hamiltonian
(\ref{1.14}) can be recast to the form
\begin{eqnarray}\label{2.8}
&&H_1(\theta,\phi(t))=-\hbar\omega\cos\theta[\cos\theta\cdot(\hat{n}_1+\hat{n}_2-1)\nonumber\\
&&+\sin\theta(e^{i\phi(t)}S^+ + e^{-i\phi(t)}S^-)],
\end{eqnarray}
or
\begin{equation}\label{2.9}
H_1(\theta,\phi(t))=-\hbar\omega\varepsilon(\theta)H_0(\theta,\phi(t))
\end{equation}
where
\begin{equation}\label{2.10}
H_0(\theta,\phi(t))=2\varepsilon(\theta)S^3+\Delta(t)S^++\Delta(t)^\ast
S^-
\end{equation}
\begin{equation}\label{2.11}
\varepsilon(\theta)=\cos\theta,\ \ \ \Delta(t)=\sin\theta
e^{i\phi(t)}
\end{equation}
The standard procedure of diagonalizing $H_0(\theta,\phi(t))$ is
\begin{equation}\label{2.12}
W^{\dag}H_0W=2{\cal E} S_3,\ \ \ {\cal
E}=\sqrt{(\varepsilon(\theta))^2+|\Delta(t)|^2}
\end{equation}
and the eigenstate is
\begin{equation}\label{2.13}
\begin{array}{ll}
&|\xi(\theta)\rangle=W|\textrm{vacuum}\rangle=\exp(\xi S_+-\xi^\ast
S_-)|\textrm{vacuum}\rangle,\\
&S_-|\textrm{vacuum}\rangle=0,
\end{array}
\end{equation}
with
\begin{equation}\label{2.14}
\xi=re^{i\phi(t)},\ \
\cot(2r)=-\frac{\varepsilon(\theta)}{|\Delta(t)|}
\end{equation}
Substituting Eq. (\ref{2.11}) into Eq. (\ref{2.12}) and Eq.
(\ref{2.14}) we obtain
\begin{displaymath}
{\cal E}=1, \ \ \ r=-\frac{\theta}{2},
\end{displaymath}
in other words, we have
\begin{equation}\label{2.15}
\begin{array}{rcl}
W^{\dag}HW|\xi(\theta)\rangle&=&-\hbar\omega\cdot 2\cos\theta
S_3|\xi(\theta)\rangle\\
&=&-\hbar\omega\cos\theta(\hat{n}_1+\hat{n}_2-1)|\xi(\theta)\rangle.
\end{array}
\end{equation}
It is nothing but an oscillator Hamiltonian formed by two fermions
with the frequency $\omega\cos\theta$. When $\theta=0$ Eq.
(\ref{2.10}) reduces to the standard oscillator for $\Delta(t)=0$.
However, When $\theta\neq0$, $\Delta(t)$ plays a role of the
``energy gap" and the wave function takes the form of spin coherent
state \cite{Gilmore}\cite{87CSS}. We know that the eigenfunction of oscillator is the Hermitian
polynomial, whereas the wave function of Eq. (\ref{2.10}) with
$\Delta(t)\neq0$ is the spin coherent state shown by
\begin{eqnarray}\label{2.16}
|\xi\rangle & = & \frac{1}{(1+|\tau|^2)^{\frac{1}{2}}} \exp(\tau
S_+)|\textrm{vacuum}\rangle\nonumber\\
& = & \frac{1}{\sqrt{1+|\tau|^2}}\{|0,0\rangle+\tau|1,1\rangle\},
\end{eqnarray}
where
\begin{eqnarray}\label{2.17}
&&\tau=-e^{i\phi}\tan\frac{\theta}{2},\nonumber\\
&&|1,1\rangle=|n_1=1, n_2=1\rangle,\nonumber\\ &&
|0,0\rangle=|n_1=0, n_2=0\rangle.
\end{eqnarray}
We thus conclude that $\theta=0$ is a point of quantum criticality.
It is not caused by temperature, but by the degree of entanglement
related to the parameter $\theta$.

The degree of entanglement (or the concurrence \cite{98Wo}) for an
arbitrary two-qubit state $|\psi\rangle = a
|\uparrow\uparrow\rangle+ b |\uparrow\downarrow\rangle+ c
|\downarrow\uparrow\rangle+ d|\downarrow\downarrow\rangle$ is ${\cal
C}=2|ad-bc|$. The Berry phases in Eq. (\ref{1.20}) can be expressed
in terms of the concurrence of the states
$|\Phi_{\pm}(\theta,\phi)\rangle$ as
\begin{eqnarray}\label{2.17a}
\gamma_{\pm}^1=\mp \pi(1-\sqrt{1-{\cal C}^2}),
\end{eqnarray}
where ${\cal C}=|\sin\theta|$ is the concurrence of
$|\Phi_{\pm}(\theta,\phi)\rangle$. Interestingly, one may observe
that when $\theta=0$ or ${\cal C}=0$, the quantum criticality occurs
in the Hamiltonian system $H_1(\theta,\phi(t))$ and at the same time
Berry phases vanish correspondingly.

\section{BP and QC for Hamiltonian $H_2(\theta,\phi(t))$}
In this section, we come to study the Berry phase and also the
quantum criticality for a kind of Yang--Baxter Hamiltonian related
to the well-known six-vertex model \cite{Jimbo1} and the
Temperley-Lieb algebra.

For the well-known six-vertex model, the braiding matrix reads

\begin{equation}\label{1.21}
   \begin{array}{lll} B&=&S^{\frac{1}{2}\frac{1}{2}}
 =
   \left[ \begin{array}{cccc}
   q&0&0&0\\
   0&0&{-\eta}&0\\
   0&{-\eta}&{q-q^{-1}}&0\\
   0&0&0&q
   \end{array} \right]\\
& =&q(I-q^{-1}U^{\frac{1}{2}\frac{1}{2}}),
   \end{array}
\end{equation}
 where
\begin{equation}\label{1.22}
   U^{\frac{1}{2}\frac{1}{2}}=
   \left( \begin{array}{cccc}
   0&0&0&0\\
   0&q&\eta&0\\
   0&\eta^{-1}&q^{-1}&0\\
   0&0&0&0
   \end{array} \right).
\end{equation}
The matrix $U^{\frac{1}{2}\frac{1}{2}}$
 satisfies the Temperley-Lieb algebra, i.e.,
$U_i U_{i\pm 1}U_i=U_i$, $U_i^2=d \ U_i$ (for the above matrix
$U^{\frac{1}{2}\frac{1}{2}}$, $d=q+q^{-1}$). The above braiding
matrix $B$ has two distinct non-zroe eigenvalues with $\lambda_1=q$,
$\lambda_2=-q^{-1}$ and $\lambda_1 \lambda_2=-1$, so we can perform
the trigonometric Yang-Baxterization approach. It gives
\begin{eqnarray}\label{1.23}
\breve{R}(x)&=&[q^2+q^{-2} -(x^2+x^{-2})]^{-1/2}[(q x -q^{-1} x^{-1})I \nonumber\\
&&-(x-x^{-1})U^{\frac{1}{2}\frac{1}{2}}],
\end{eqnarray}
\begin{eqnarray}\label{1.24}
[\breve{R}(x)]^{-1}&=&[q^2+q^{-2} -(x^2+x^{-2})]^{-1/2}[(q x^{-1} -q^{-1} x)I \nonumber\\
&&+(x-x^{-1})U^{\frac{1}{2}\frac{1}{2}}].
\end{eqnarray}
It is easy to check that
$[\breve{R}(x)]^\dag$=$[\breve{R}(x)]^{-1}$=$\breve{R}(-x)$ for
$x=e^{i\vartheta}$, $\eta=e^{i\varphi(t)}$, and $\theta,
\varphi(t), q\in \textrm{real}$.

One may symmetrize the matrix $\breve{R}(x)$ given by Eq. (\ref{1.23})
(i.e., to make the matrix elements
$\left[\breve{R}(x)\right]^{1/2,-1/2}_{1/2,-1/2}
=\left[\breve{R}(x)\right]^{-1/2,1/2}_{-1/2,1/2}$ ) through the
following unitary transformation
\begin{equation}\label{1.25}
\breve{R}_{i\ i+1}(V(x))=V(x)\breve{R}_{i\ i+1}(x) V(x)^\dag,
\end{equation}
where $V(x)=V_i(x)\otimes [V_{i+1}(x)]^{-1}$ and
\begin{equation}\label{1.26}
V_i(x)=\left(\begin{array}{cc} 0 & x^{-\frac{1}{4}}\\
x^{\frac{1}{4}} & 0
\end{array}\right).
\end{equation}
The resultant $\breve{R}_{i\ i+1}(V(x))$ is still a solution of YBE.
Let only the parameter $\eta=e^{i\varphi(t)}$ be time-dependent, it
yields from Eq. (\ref{1.5}) and Eq. (\ref{1.25}) that
\begin{equation}\label{1.27}
\begin{array}{lll}
H_2(x, \varphi(t))&=&\hbar
\dot{\varphi}\left[q^2+q^{-2}-(x^2+x^{-2})\right]^{-1}(x-x^{-1})\\
&&\times
[(x-x^{-1})(S_1^3-S_2^3)+\\&&(q-q^{-1})(e^{i\varphi}S_1^+S_2^- -
e^{-i\varphi}S_1^-S_2^+)].
\end{array}
\end{equation}
Putting $x=e^{i\vartheta}$, $\vartheta=\pi/2-\theta$ and
$\varphi(t)=\phi(t)-\pi/2=\omega t$, we have
\begin{equation}\label{1.28}
\begin{array}{lll}
H_2(\theta, \phi(t))&=&-4\hbar\omega \left[q^2+q^{-2}+2\cos 2\theta
\right]^{-1}\cos \theta  \\
&&\times [\cos \theta(S_1^3-S_2^3)+\\
&&\frac{1}{2}(q-q^{-1})(e^{i\phi} S_1^+S_2^-+e^{-i\phi}S_1^-S_2^+)],
\end{array}
\end{equation}
whose two nonzero eigenvalues are
\begin{eqnarray}\label{1.29}
E_{\pm}^2&=&-4\hbar\dot{\phi}(q^2+q^{-2}+2\cos
2\theta)^{-1}\cos\theta
\lambda_{\pm}\nonumber\\&=&-\frac{4\hbar\dot{\phi}
\cos\theta}{\lambda_{\pm}},
\end{eqnarray}
with
\begin{equation}\label{1.30}
\lambda_{\pm}=\pm\sqrt{\cos^2\theta+(q-q^{-1})^2/4}.
\end{equation}
Under the adiabatic approximation the corresponding eigenstates are
\begin{equation}\label{1.31}
\begin{array}{rcl}
\left|\Phi_+(\theta,\phi)\right>&=&\frac{1}{\sqrt{2\lambda_+}}[(\lambda_+-\cos\theta)^{
-1/2}\left(\frac{q-q^{-1}}{2}\right)\left|\uparrow\downarrow\right>\\
&&+ i(\lambda_+-\cos \theta)^{1/2}e^{-i\phi}\left|\downarrow\uparrow
\right>],\\
\left|\Phi_-(\theta,\phi)\right>&=&\frac{1}{\sqrt{2\lambda_-}}[i(\lambda_--\cos\theta)^{
-1/2}\left(\frac{q-q^{-1}}{2}\right)e^{i\phi}
\left|\uparrow\downarrow\right> \\
&&-(\lambda_--\cos \theta)^{1/2}e^{-i\phi}\left|\downarrow\uparrow
\right>].
\end{array}
\end{equation}
The corresponding Berry phases for the Yang--Baxter system are
\begin{eqnarray}\label{1.32}
\gamma_{\pm}^2 &=
&\pm\pi\biggr(1-\frac{1}{\lambda_+}\cos\theta\biggr)\nonumber\\ & =&
\pm\pi\biggr[1-\frac{\cos\theta}{[\cos^2\theta+(q-q^{-1})^2/4]^{1/2}}\biggr].
\end{eqnarray}
The above Berry phases have been ``$q$-deformed", when
$\lambda_+=1$, or $q=\sqrt{1+\sin^2\theta}\pm \sin\theta$,
 Eq. (\ref{1.32}) reduces to Eq. (\ref{1.20}). Remarkably the Berry
 phases in Eq. (\ref{1.32}) can
still be expressed in terms of the concurrence of the states
$|\Phi_{\pm}(\theta,\varphi)\rangle$ in Eq. (\ref{1.31}) as
$\gamma_{\pm}^2=\mp \pi(1-\sqrt{1-{\cal C}^2})$, where ${\cal
C}=(q-q^{-1})/(2 \lambda_+)$.

Similarly, the Hamiltonian $H_2(\theta,\phi(t))$ can be rewritten in
terms of $SU(2)$ generators $J^+=S_1^+S_2^-=\hat{f}_1^\dag
\hat{f}_2$, $J^-=S_1^-S_2^+=\hat{f}_1 \hat{f}_2^\dag$,
$J^3=(S_1^3-S_2^3)=(\hat{n}_1-\hat{n}_2)/2$ as
\begin{equation}\label{2.9a}
H_2(\theta,\phi(t))=-4\hbar\omega \frac{\cos
\theta}{q^2+q^{-2}+2\cos 2\theta} H'_0(\theta,\phi(t)),
\end{equation}
where
\begin{eqnarray}\label{2.10a}
&&H'_0(\theta,\phi(t))=2\varepsilon(\theta)J^3+\Delta(t)J^++\Delta(t)^\ast
J^-, \nonumber\\
&& \varepsilon(\theta)=\cos\theta,\ \ \ \Delta(t)=
e^{i\phi(t)}(q-q^{-1})/2.
\end{eqnarray}
When $q-q^{-1}=0$, or $q=\pm 1$, the Hamiltonian
$H'_0(\theta,\phi(t))$ contracts to
$H'_0(\theta,\phi(t))=\varepsilon(\theta)(\hat{n}_1-\hat{n}_2)$,
thus the quantum criticality occurs. Correspondingly, one may easily
see that the Berry phases in Eq. (\ref{1.32}) vanish.

\section{Conclusion and Discussion}

In summary, we have obtained some spin-1/2 interaction Hamiltonians
from the unitary Yang--Baxter $\breve{R}$-matrix. Based on which,
Berry phases and quantum criticality in the Yang--Baxter systems
have been studied.

Let us make three discussions to end this paper.

(i) In Sec. II and Sec. III, we have focused on the trigonometric
Yang-Baxterization approach
$B(\phi)\rightarrow\breve{R}(\theta,\phi)$, based on which the
Yang--Baxter Hamiltonians $H_1(\theta,\phi(t))$ and
$H_2(\theta,\phi(t))$ have been established. Now let us come to
discuss another approach called the rational Yang-Baxterization
\cite{GeXue}. Actually, the first $\breve{R}(u)$-matrix discovered
in Ref. \cite{YangCN} is a good and simple example for the rational
Yang-Baxterization approach. The $\breve{R}(u)$-matrix reads
\begin{equation}\label{RYang}
\breve{R}_{i,i+1}(u)=\frac{-c+iuP_{i,i+1}}{c+iu},
\end{equation}
where $c$ is a parameter appeared in the one-dimensional
$\delta$-function interaction Hamiltonian:
$H=-\sum_{i=1}^{N}\frac{\partial^2}{\partial x_i^2}+
2c\sum_{i<j}\delta(x_i-x_j)$, and $u=k_i-k_{i+1}$ is the relative
momentum between the $i$-th particle and the $(i+1)$-th particle. If
$c=0$, the Hamiltonian represents $N$ free particles without any
interaction, and correspondingly $\breve{R}_{i,i+1}(u)$-matrix
reduces to $P_{i,i+1}$. Generally, for a given braiding matrix
$B_{i,i+1}$, we may perform the following transformation
\begin{equation}\label{RYang1}
\breve{R}_{i,i+1}(u)=\rho(u) \; \frac{\delta+ \gamma  u
B_{i,i+1}}{\alpha+\beta u},
\end{equation}
if the $\breve{R}_{i,i+1}$ matrices obey the YBE, then we call Eq.
(\ref{RYang1}) as the rational Yang-Baxterization approach.

The Yang--Baxter Hamiltonians are induced from the
$\breve{R}(\theta,\phi)$-matrix via Eq. (\ref{1.5}). It is natural
to ask whether the same Hamiltonian, e.g., $H_1(\theta,\phi(t))$,
can be induced from different matrices of $\breve{R}(\theta,\phi)$.
The answer is yes. Here we would like to provide such an example.

Considering the following braiding matrix
\begin{equation}\label{1.34}
B=S^{\frac{1}{2}\frac{1}{2}}=I+fU^{\frac{1}{2}\frac{1}{2}},
\end{equation}
where
\begin{equation}\label{1.35}
U^{\frac{1}{2}\frac{1}{2}}=\left(
\begin{array}{cccc}
\epsilon&0&0&e^{i\varphi}\\
0&0&0&0\\
0&0&0&0\\
e^{-i\varphi}&0&0&\epsilon
\end{array}
\right)
\end{equation}
satisfies the TL algebra, and $d=2\epsilon$, $\beta=-d/2=-\epsilon$,
$\epsilon=\pm 1$, $f=(-d\pm\sqrt{d^2-4})/2=\beta$. After performing
the rational Yang--Baxterization, from Eqs. (\ref{RYang1}) and
(\ref{1.34}) one obtains the $\breve{R}(u)$ satisfying the YBE as
\begin{equation}\label{1.33}
\breve{R}(u)=I+G(u)U, \ \ \ G(u)=\frac{u}{\alpha+\beta u}.
\end{equation}
Furthermore, the unitary condition
$[\breve{R}(u)]^{\dag}=[\breve{R}(u)]^{-1}=\breve{R}(-u)$ leads to
$G(-u)=G(u)^*$, or $(\alpha^{-1}u)^* = - \alpha^{-1} u$. We choose
\begin{equation}\label{1.39}
\alpha u^{-1} = i \tan \theta,
\end{equation}
then it is easy to have from Eq. (\ref{1.5}) and Eq. (\ref{1.33})
that
\begin{equation}\label{1.36}
H(u,\phi)=i \hbar G(u) G(-u) \frac{\partial U}{\partial
t}[G(-u)^{-1}I+U].
\end{equation}
Substituting Eq. (\ref{1.33}) into Eq. (\ref{1.36}) one obtains
\begin{eqnarray}\label{1.40}
H_3(\theta,\varphi(t))&=& - \hbar \dot{\varphi} \cos \theta \{\cos
\theta
(S_1^3 +S_2^3)\nonumber\\
&& - i \sin \theta [e^{i \varphi}S_1^{+}S_2^{+}-e^{-i
\varphi}S_1^{-}S_2^{-}]\}.
\end{eqnarray}
After redefining the parameter $\varphi(t)=\phi(t)-\pi/2$, one may
find that the Hamiltonian $H_3(\theta,\varphi(t))$ is identical to
the Hamiltonian $H_1(\theta,\phi(t))$ as shown in Eq. (\ref{1.14}).

(ii) The same matrix $U^{\frac{1}{2}\frac{1}{2}}$ may yield
inequivalent Yang--Baxter Hamiltonians. For instance, Let
\begin{equation}\label{1.35a}
U^{\frac{1}{2}\frac{1}{2}}=\left(
\begin{array}{cccc}
0&0&0&0\\
0&\epsilon&e^{i\varphi}&0\\
0&e^{-i\varphi}&\epsilon&0\\
0&0&0&0
\end{array}
\right),
\end{equation}
which is a special form of Eq. (\ref{1.22}) by taking
$q=\epsilon=\pm 1$. Based on the same matrix
$U^{\frac{1}{2}\frac{1}{2}}$, one can have two kinds of inequivalent
braiding matrices, one is
$B_1=q(I-q^{-1}U^{\frac{1}{2}\frac{1}{2}})$ in Eq. (\ref{1.21}), the
other is $B_2=I+fU^{\frac{1}{2}\frac{1}{2}}$ in Eq.
(\ref{1.34}). After making the trigonometric Yang-Baxterization for
the former one $B_1$, it yields the Yang--Baxter Hamiltonian
$H_2(\theta,\phi(t))$ as in Eq. (\ref{1.28}); similarly, after
making the rational Yang-Baxterization for the latter one $B_2$, it
yields the following Yang--Baxter Hamiltonian:
\begin{eqnarray}\label{1.40a}
H_4(\theta,\phi(t))&=& - \hbar \dot{\phi} \cos \theta \{\cos \theta
(S_1^3 -S_2^3)\nonumber\\
&& + \sin \theta [e^{i \phi}S_1^{+}S_2^{-}+e^{-i
\phi}S_1^{-}S_2^{+}]\}.
\end{eqnarray}
The Hamiltonian $H_4(\theta,\phi(t))$ is inequivalent to Hamiltonian
$H_2(\theta,\phi(t))$, and it has the same eigenvalues and Berry
phases as those of the Hamiltonian $H_1(\theta,\phi(t))$.

(iii) For the spins at $N$-lattices one may define the following
$SU(2)$ generators as
\begin{eqnarray*}
S^+=\prod^N_{i=1}S^{+}_i=\prod^N_{i=1}\hat{f}^{\dag}_i=
\left(\begin{array}{ccccc}
 & & & & 1\\
 & & & 0 & \\
 & & \cdots& &\\
  & 0& &  & \\
 0  & & & &
\end{array}\right),
\end{eqnarray*}
\begin{eqnarray*}
S^-=\prod^N_{i=1}S^{-}_i=\prod^N_{i=1}\hat{f}_i=
\left(\begin{array}{ccccc}
 & & & & 0\\
 & & & 0 & \\
 & & \cdots& &\\
  & 0& &  & \\
 1  & & & &
\end{array}\right),
\end{eqnarray*}
\begin{eqnarray}
S^3&=&[S^+, S^-]/2=\biggr(\prod^N_{i=1}\hat{n}_i -
\prod^N_{i=1}(1-\hat{n}_i)\biggr)/2\nonumber\\
&=& \frac{1}{2}\left(\begin{array}{ccccc}
 1& & & & \\
 & 0& &  & \\
 & & \ddots& &\\
  & & &  & \\
   & & & & -1
\end{array}\right).
\end{eqnarray}
Similarly, the $SU(2)$ Casimir operator is $\frac{1}{2}(S^+S^-
+S^-S^+)+(S^3)^2=\frac{3}{4}(\prod^N_{i=1}\hat{n}_i +
\prod^N_{i=1}(1-\hat{n}_i))$, which equals to
$\frac{1}{2}(\frac{1}{2} + 1)$ if all $\hat{n}_i$'s are equal, and
otherwise $0(0+1)$.

When $S^+$ acts on the vacuum state, it produces the state with all
spins up $|\uparrow\uparrow\cdots \uparrow\rangle$, similarly, $S^-$
produces the state with all spins down $|\downarrow\downarrow\cdots
\downarrow\rangle$. The states $|\uparrow\uparrow\cdots
\uparrow\rangle$ and $|\downarrow\downarrow\cdots \downarrow\rangle$
are the chiral spin states, or the ``chiral photons". All the
similar discussion on quantum criticality and Berry phase can be
extended to multipartite spin-1/2 systems. Eventually, people have
currently found that braiding operators have some important physical
applications in non-Abelian quantum Hall systems, topological
quantum field theory and topological quantum computation
\cite{Wilczek0,Slingerland,Kitaev,Bonesteel}, how to apply the
Yang--Baxter $\breve{R}(\theta,\phi)$-matrix (that is a
generalization of the braiding operator) to these fields is an
interesting and significant topic. We shall investigate this subject
subsequently.


{\bf ACKNOWLEDGMENTS}  The authors thank Prof. J. L. Birman for his
useful discussion. This work is supported in part by NSF of China
(Grants No. 10575053 and No. 10605013) and Program for New Century
Excellent Talents in University. The Project-sponsored by SRF for
ROCS, SEM.

\end{document}